\documentclass {article}
\usepackage{spconf,amsmath,graphicx}
\usepackage{algorithm}
\usepackage{algorithmic}
\usepackage{multirow}

\newcommand{\tabincell}[2]{\begin{tabular}{@{}#1@{}}#2\end{tabular}}  
\newcommand\ChangeRT[1]{\noalign{\hrule height #1}}

\title{Architecture-aware Network Pruning for vision quality applications}

\name{Wei-Ting Wang, Han-Lin Li, Wei-Shiang Lin, Cheng-Ming Chiang, Yi-Min Tsai\thanks{Copyright 2019 IEEE. Published in the IEEE 2019 International Conference on Image Processing (ICIP 2019), scheduled for 22-25 September 2019 in Taipei, Taiwan. Personal use of this material is permitted. However, permission to reprint/republish this material for advertising or promotional purposes or for creating new collective works for resale or redistribution to servers or lists, or to reuse any copyrighted component of this work in other works, must be obtained from the IEEE. Contact: Manager, Copyrights and Permissions / IEEE Service Center / 445 Hoes Lane / P.O. Box 1331 / Piscataway, NJ 08855-1331, USA. Telephone: + Intl. 908-562-3966.
}}
\address{MediaTek Inc.}

\begin{document}
\maketitle
\normalsize

\begin{abstract}
Convolutional neural network (CNN) delivers impressive
achievements in computer vision and machine learning field. 
However, CNN incurs high computational complexity, 
especially for vision quality applications
because of large image resolution. 
In this paper, we propose an iterative architecture-aware pruning algorithm with adaptive magnitude threshold while cooperating with quality-metric measurement simultaneously.
We show the performance improvement applied on vision quality applications and
provide comprehensive analysis with flexible pruning configuration.
With the proposed method, the Multiply-Accumulate (MAC) of state-of-the-art low-light imaging (SID) and super-resolution (EDSR) are reduced by 58\% and 37\% without quality drop, 
respectively. The memory bandwidth (BW) requirements of convolutional layer can be also reduced by 20\% to 40\%.
\end{abstract}
\section{Introduction}
\label{sec:intro}
CNN is adopted as an essential ingredients in computer vision and machine learning areas 
\cite{classfication,detection,segmentation}.
 Vision perception tasks including image classification, object detection and semantic segmentation are comprehensively investigated and associated with CNN. 
Even in image processing field such as super resolution, high dynamic range imaging and de-noising, 
CNN has progressive and promising improvement on image quality in recent years \cite{sid,edsr}.

However, compared to perception tasks, it requires higher computational complexity and BW requirements for vision quality tasks.
MobileNetV1 \cite{mobilenet} is designed with 569M MAC for ImageNet classification. 
On the other hand, in low-light photography SID \cite{sid} and super-resolution EDSR \cite{edsr}, 
it takes 560G MAC and 1.4T MAC per inference, respectively. 
It is more challenge to deploy CNN models on mobile devices for vision quality applications.

\begin{figure}
\begin{minipage}[t]{0.45\linewidth}
  \centering
  \centerline{\includegraphics[width=4.4cm]{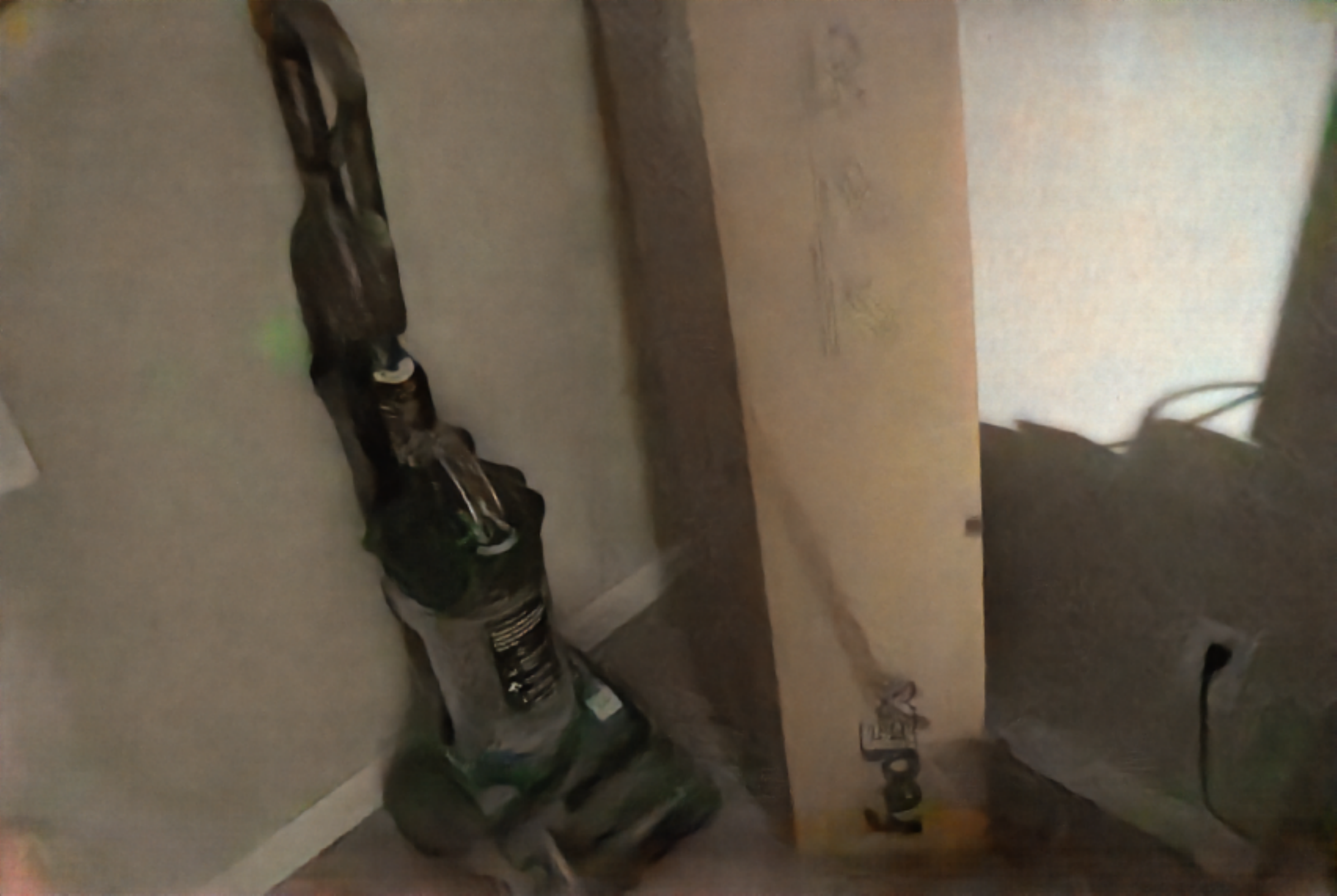}}
  \begin{center}
    (a)
  \end{center}
\end{minipage}
\hspace{0.06\linewidth}
\begin{minipage}[t]{0.45\linewidth}
  \centering
  \centerline{\includegraphics[width=4.4cm]{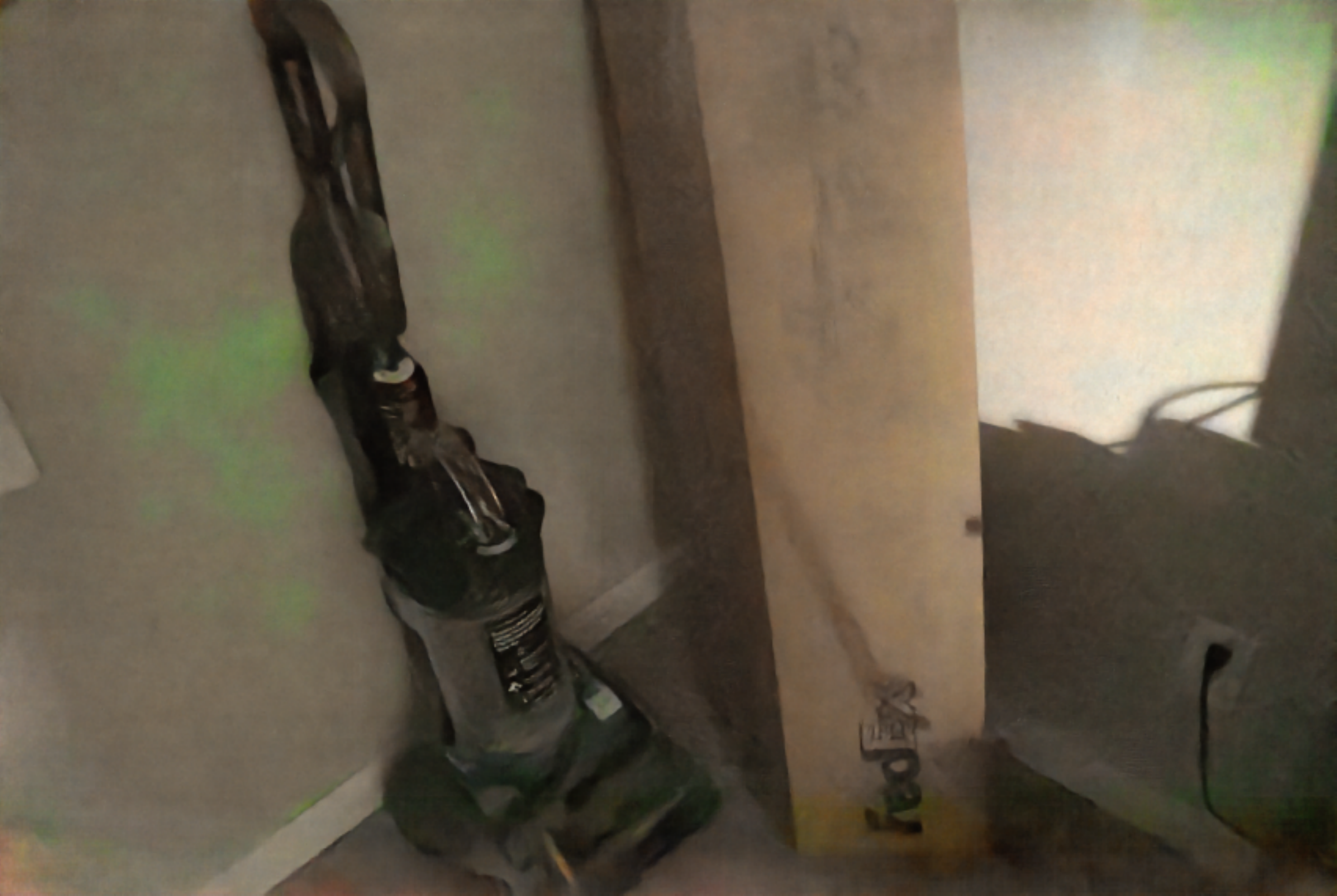}}
  \begin{center}
  	(b) 
  \end{center}
\end{minipage}
\caption{Slightly quality-metric drop (PSNR -0.09) may incur visible defects (SID). (a) PSNR: 25.41. (b) PSNR: 25.32.}
\label{fig:psnrdrop}
\end{figure}

Network pruning \cite{Pruning} is an effective methodology toward performance optimization.
Sparsity is defined as the ratio of the number of zero weights divided by the number of total weights. 
Better pruning algorithm delivers higher sparsity 
and reduces more MAC and BW correspondingly. 
However, quality drop is one of the major challenges in network pruning.
Fig. \ref{fig:psnrdrop} shows visible defect on SID even with only 0.1 PSNR degradation.

In this paper, we propose architecture-aware pruning to maximize sparsity and MAC reduction without quality-metric (PSNR or SSIM) drops. We also analyze the effects of MAC and BW reduction with different configurations associating with pruned structures. The proposed method focus on algorithms including but not limited to SID and EDSR.

\section{Related Works }
\label{sec:relate_work}

Network pruning has been widely explored in existing literatures.
To answer which weight should be pruned, some works add evaluation functions to loss function, such as group lasso \cite{ssl} and MAC regularization \cite{MorphNet}. However, it is difficult to find a proper ratio between additional pruning-related loss and original loss. 
Others works create evaluation functions, including sensitivity \cite{optimal_brain_damge,Engelbrecht:2001:NPH:2325782.2326806} and weight magnitude \cite{han2015}.
The sensitivity method computes the impact of weights on the training loss and removes low-impact weights. Weight magnitude method simply prune weight if its absolute value is less than the threshold, which is easier to be applied on large-scale CNNs. In this work, we use weight magnitude method to prune the network.\\

{\setlength{\parindent}{0cm}\textbf{Pruning granularity.} There are two granularity of pruning, fine-grained pruning \cite{han2015, finepruning1} and coarse-grained pruning
\cite{coarsepruning1, coarsepruning2, coarsepruning3, coarsepruning4}. 
Fined-grained method prunes individual weights (i.e., within a filter kernel), whereas coarse-grained method extensively considers network structures (i.e., along the output and the input channels). 
According to \cite{AMC}, fine-grained pruning 
needs additional dedicated hardware to handle irregular sparsity.
Coarse-grained method may obtain higher compression ratio 
without the need of compression header \cite{han2017}.
Therefore, we focus on coarse-grained output-channel-wise pruning.}\\

{\setlength{\parindent}{0cm}\textbf{Iterative pruning}. To prevent catastrophic accuracy degradation, iterative pruning is viewed as an effective retraining procedure \cite{DBLP:journals/corr/YangCS16a, DBLP:journals/corr/HuPTT16, iterative1}.
For vision quality applications, quality metrics are required as a reference judgement for termination of pruning procedure.}

\section{PROPOSED METHOD}
\label{sec:method}
\subsection{Architecture-aware Pruning}
An output channel is pruned if its maximum absolute weight value is less than magnitude threshold.
For convolutional layer, the weight kernel has tensor shape 
$i\times o\times k\times k$, where 
$i$ is the number of input channels,
$o$ is the number of output channels and
$k$ is kernel size.
Output-channel-wise pruning removes 
the weights along output channels.
The kernel shape becomes $i\times (o-o')\times k\times k$ if $o'$ output channels are pruned.
The output-channel pruned ratio is defined as $o'/o$.

Once output channels of a layer are pruned,
the corresponding input channels of the 
following layer are also removed.
We defined one layer's sparsity as 
$1-((i-i')(o-o'))/(i\times o)$, 
where $i'$ is the number of pruned input channel in the layer.
The network sparsity is defined as the ratio of the number of zero weights of a pruned network divided by the number of total weights of the original network.

\subsubsection{Keep Layer Depth}
\label{ssec:keepnetworktopology}
Usually, in vision quality applications, each layer in network is semantically 
designed for quality-sensitive primitives, 
such as edge and chroma, with respect to different resolutions. 
Intensively removing a layer can severely degrade the quality. 
Therefore, we keep the network architecture
by preserving minimum number of output channels.

\subsubsection{Enhance MAC Efficiency}
\label{ssec:macefficient}
\begin{figure}[t]
\begin{minipage}[b]{1\linewidth}
  \centering
  \centerline{\includegraphics[width=8cm]
  {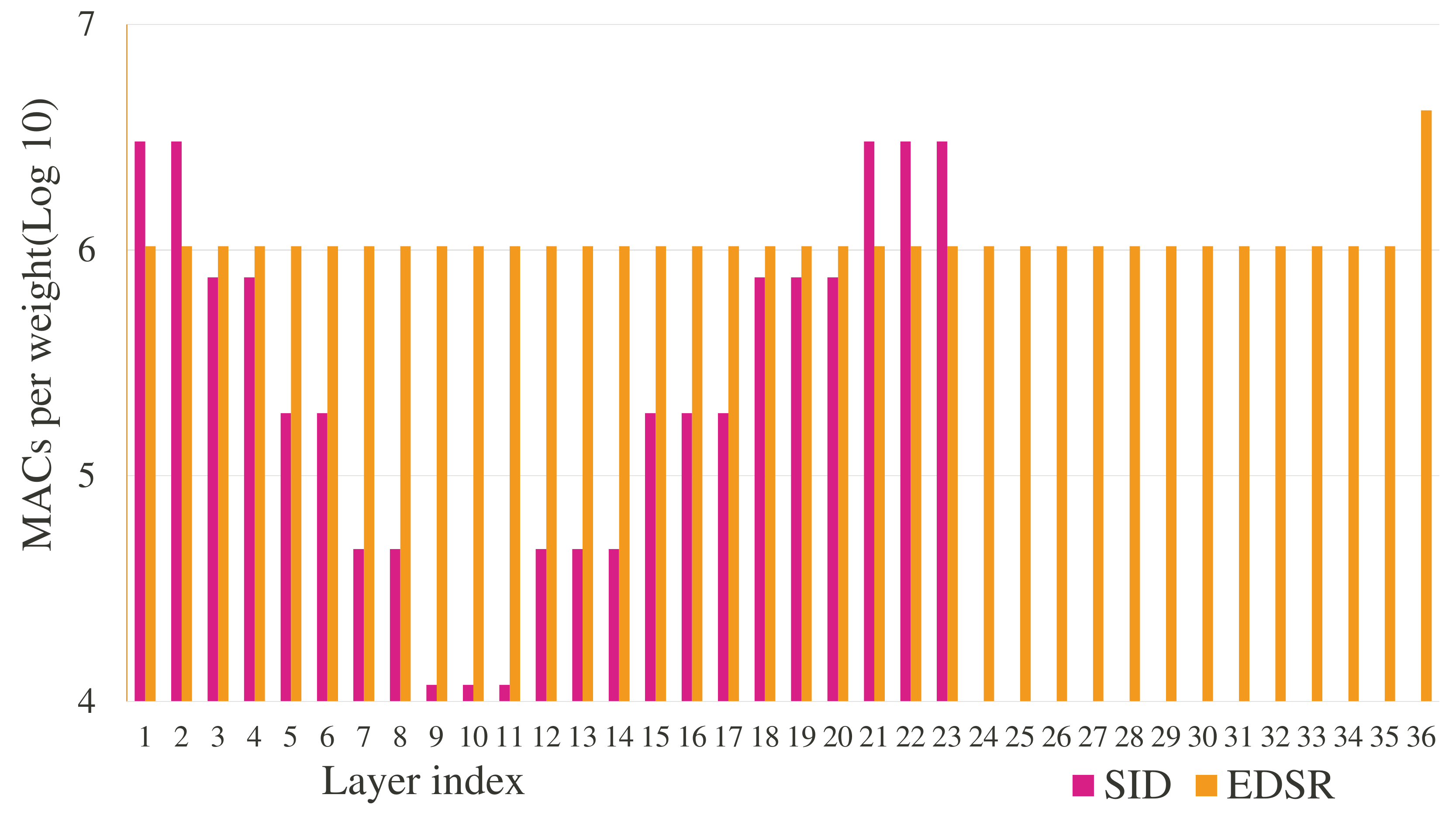}}
  \medskip
\end{minipage}
\caption{MAC/weight are much larger on top and bottom layers in SID. However, MAC/weight for most layers are uniform in EDSR.}
\label{fig:macperweight}
\end{figure}

A pruned network with higher weight sparsity 
may not imply higher computation reduction. 
We define MAC/weight, $R_l$ (Eq. \ref{equ:macefficient}), for each layer 
as an indicator of MAC efficiency. $M_l$ and $W_l$ is number of MAC and weights of a layer, respectively.
Fig. \ref{fig:macperweight} shows that MAC/weight are much larger on both top and bottom layers on SID because of its U-Net \cite{unet} network topology.
Therefore, to productively reduce computation, output channels of a layer with higher $R_l$ are tend to be pruned more because of higher magnitude threshold.

\begin{equation}
\label{equ:macefficient}
 R_l = log_{10}(M_l/W_l) 
\end{equation} 

\begin{figure}[t]
\begin{minipage}[b]{1\linewidth}
  \centering
  \centerline{\includegraphics[width=6cm]
  {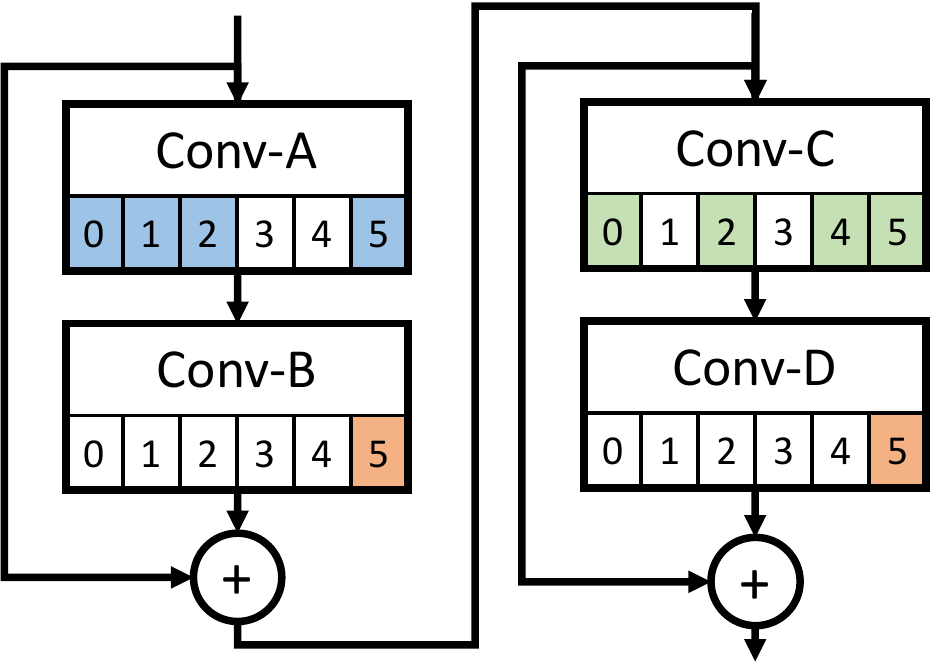}}
  \medskip
\end{minipage}
\caption{An example of residual block. There are 4 layers with 6 output channels in each layer. Color parts represent removed output channels. One color denotes one group of channels in balance pruned output channel method.}
\label{fig:residualblock}
\end{figure}

\subsubsection{Balance Pruned Output Channel}
\label{ssec:shortcutblocks}
Residual block is universally used in network topology design such as in EDSR which is a variation of ResNet \cite{resnet} with long shortcut. 
However, to prune output channels from residual block is arduous because of element-wise operations (i.e., element-wise ADD) or concatenation. 
Fig. \ref{fig:residualblock} illustrates an example that a magnitude threshold is applied to 4-layer residual block. 
Because of element-wise ADD after layer Conv-B and layer Conv-D, the output channel of a given layer (Conv-D) and its preceding layer (Conv-B) with the same index (5) should be grouped and pruned at the same time.
Conv-B layer and Conv-D layer have less pruned output channels compared to layer Conv-A and layer Conv-C.

We propose a guidance (Eq. \ref{equ:balance_threshold}) to prune output channels 
of residual block easier by increasing magnitude threshold on layers with lower ratio of pruned output channels. 
MAC efficiency mentioned in Sec. \ref{ssec:macefficient} is also applied.
$S_l$ is the ratio of pruned output channels of layer $l$.
$T_b$ is the original magnitude threshold base. 
Thus, output channels of a layer with lower $S_l$ have higher tendency to be pruned.

\begin{equation}
\label{equ:balance_threshold}
 T_l = T_b \times (1-S_l) \times R_l
\end{equation}

\subsection{Quality Metric Guarantee}
\label{ssec:qualitymetricguarantee}
To maintain quality metric (PSNR and SSIM)
while maximizing pruned MAC,
our algorithm prunes and retrains network iteratively. The iteration terminates when either the target quality-metric criteria or maximum training steps is reached.
The proposed overall flow is shown in Algorithm \ref{alg:prune}.

\begin{algorithm}[t]
   \caption{Architecture-aware and Quality Metric Guaranteed Pruning}
   \label{alg:prune}
\begin{algorithmic}[1]
   \STATE {\bfseries Input:} 
   Target quality $Q$, 
   Target Sparsity Increment $S_i$,
   Threshold Increment $T_i$,
   Total Step $G$
   \STATE Target Sparsity $S = S_i + $ total-network-sparsity   
   \STATE Initial Threshold Base $T_b = T_i$
   \REPEAT
   \FOR{layer {\bfseries in} network}
   \STATE $S_l =$ pruned-output-channel-ratio(layer)
   \STATE $M_l =$ MAC(layer)
   \STATE $W_l =$ weight-size(layer)
   \STATE $R_l = log_{10}(M_l/W_l)$
   \STATE $T_l = T_b \times(1-S_l)\times R_l$
   \STATE prune-output-channels-by-threshold $T_l$
   \ENDFOR
   \STATE $S_c=$ calculate-total-network-sparsity
   \STATE $T_b = T_b + T_i$
   \UNTIL{$S_c > S$}
   \REPEAT
   \STATE retrain-pruned-network
   \STATE $Q_t =$ evaluate-quality-metric
   \STATE $g=$ get-current-step
   \UNTIL($Q_t > Q$ or $g>=G$)
   \IF {$g<G$}
   \STATE jump	 to line 2
   \ENDIF
\end{algorithmic}
\end{algorithm}

\section{Experimental Result}
\label{sec:exp}

\subsection{Experiment Setup}
We generally investigate both 
SID for low-light photography and 
EDSR (baseline network, $\times$2) for super resolution.
SID uses its own dataset \cite{sid} and EDSR adopts DIV2K dataset \cite{div2k}.
The input size of the network is set to the maximum image resolution in the datasets, 
1424$\times$2128 for SID and 1020$\times$1020 for DIV2K, to calculate MAC and BW. 

In SID dataset, we use images captured by Sony $\alpha$7SII camera as our training and validation data, which contains 280 pairs and 93 pairs, respectively. The pre-process stage is aligned with the setting in SID paper.
In DIV2K dataset, we use the pre-process setting mentioned in \cite{edsr} to generate 5,458,040 training patches from 800 training images and use 100 validation images as our validation data.

\begin{table*}
\caption{Detailed results. BW, considering only convolutional layers, consists of both weights and activations.
Each weight and activation is represented with 4-byte floating-point numerical precision.}

\label{tab:expresult}
\begin{center}
\begin{tabular}{r|l|rr|rr|rr|rr|c|c}
	\ChangeRT{1.6pt}
	\textbf{Network} & 
	\textbf{Solution} &
	\multicolumn{2}{c|}
	{\textbf{\tabincell{c}{$\#$ of MAC \\$(\times10^9)$}}} &
	\multicolumn{2}{c|}
	{\textbf{\tabincell{c}{$\#$ of Weights \\$(\times10^3)$}}} &
	\multicolumn{2}{c|}
	{\textbf{\tabincell{c}{$\#$ of Activations \\$(\times10^6)$}}} &
	\multicolumn{2}{c|}
	{\textbf{\tabincell{c}{BW \\$($MByte/Inference$)$}}} &
	\textbf{\tabincell{c}{Validation \\ PSNR}} &
	\textbf{\tabincell{c}{Validation \\ SSIM}} \\ 
	\ChangeRT{1.6pt}  
	SID    
    &\tabincell{l}{Original}
    & 560 & (100\%) & 7757 & (100\%) & 1915 & (100\%) & 1922 & (100\%) & 28.54 & 0.767    
	\\   
    SID 
    &\tabincell{l}{Method-A}
    & 458 & (82\%) & 6918 & (89\%) & 1632 & (85\%) & 1639 & (85\%) & 28.54 & 0.768     
	\\     
	SID
	&\tabincell{l}{Method-B}
    & 354 & (63\%) & 5275 & (68\%) & 1485 & (78\%) & 1491 & (78\%) & 28.54 & 0.771    
	\\     
	SID
	&\tabincell{l}{Method-C}
    & 270 & (48\%) & 5584 & (72\%) & 1219 & (64\%) & 1225 & (64\%) & 28.54 & 0.769    
	\\ 
	SID
	&\tabincell{l}{Method-D}
    & 236 & (42\%) & 4241 & (55\%) & 1169 & (61\%) & 1173 & (61\%) & 28.55 & 0.768
    \\ 
    \ChangeRT{1pt}
    EDSR
    &\tabincell{l}{Original}
    & 1428 & (100\%) & 1367 & (100\%) & 5076 & (100\%) & 5077 & (100\%) & 34.42 & 0.942
    \\ 
	EDSR    
    &\tabincell{l}{Method-A}
    & 1085 & (76\%) & 1037 & (76\%) & 4481 & (88\%) & 4481 & (88\%) & 34.43 & 0.942    
    \\ 
	EDSR
	&\tabincell{l}{Method-B}
    & 1085 & (76\%) & 1037 & (76\%) & 4481 & (88\%) & 4481 & (88\%) & 34.43 & 0.942
    \\ 
	EDSR
	&\tabincell{l}{Method-C}
    & 1085 & (76\%) & 1037 & (76\%) & 4481 & (88\%) & 4481 & (88\%) & 34.43 & 0.942    
    \\
	EDSR
	&\tabincell{l}{Method-D}
    & 897 &  (63\%) & 857 & (63\%) & 4083 & (80\%) & 4083 & (80\%) & 34.42  & 0.942 
    \\ 
	\ChangeRT{1.6pt}
    
\end{tabular}
\end{center}
\end{table*}

\begin{figure}
\begin{minipage}[t]{0.45\linewidth}
  \centering
  \centerline{\includegraphics[width=4.4cm]
  {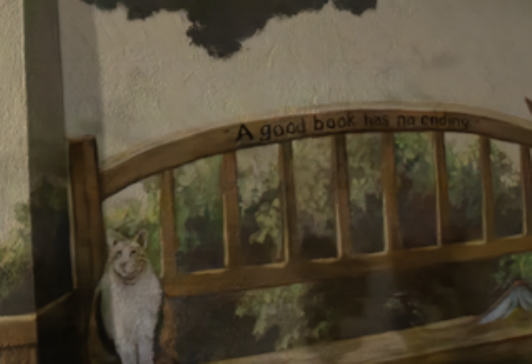}}
  \centerline{(a)}  
\end{minipage}
\hspace{0.05\linewidth}
\begin{minipage}[t]{0.45\linewidth}
  \centering
  \centerline{\includegraphics[width=4.4cm]
  {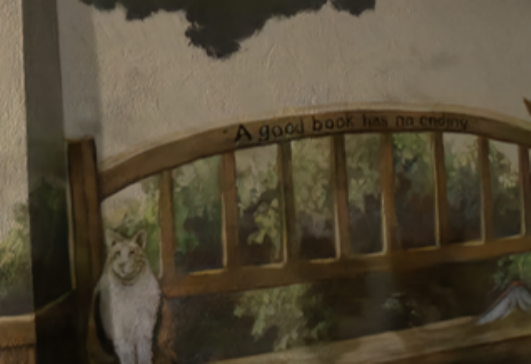}}
  \centerline{(b)}    
\end{minipage}
\begin{minipage}[t]{0.45\linewidth}
  \centering
  \centerline{\includegraphics[width=4.4cm]
  {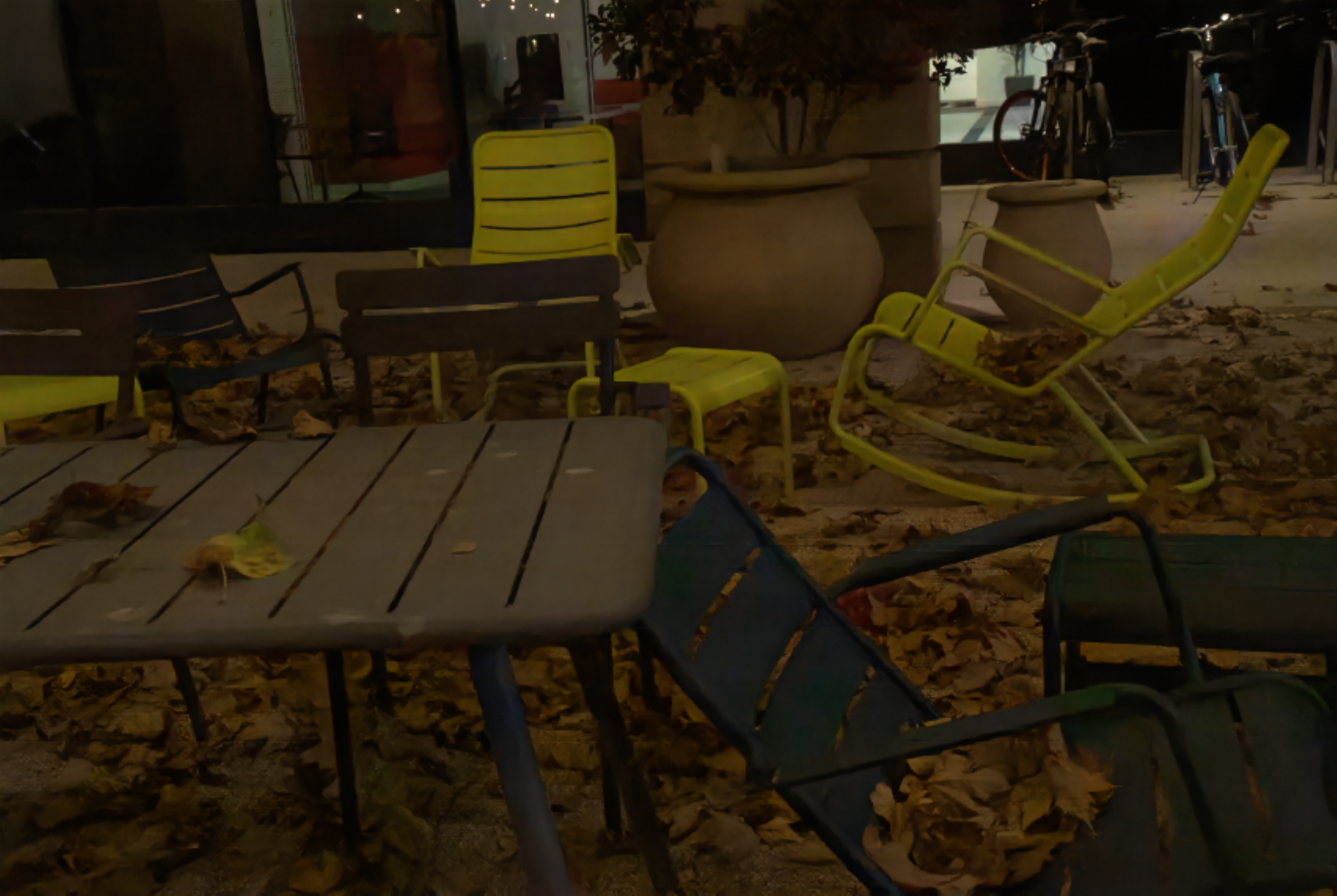}} 
  \centerline{(c)}  
\end{minipage}
\hspace{0.08\linewidth}
\begin{minipage}[t]{0.45\linewidth}
  \centering
  \centerline{\includegraphics[width=4.4cm]
  {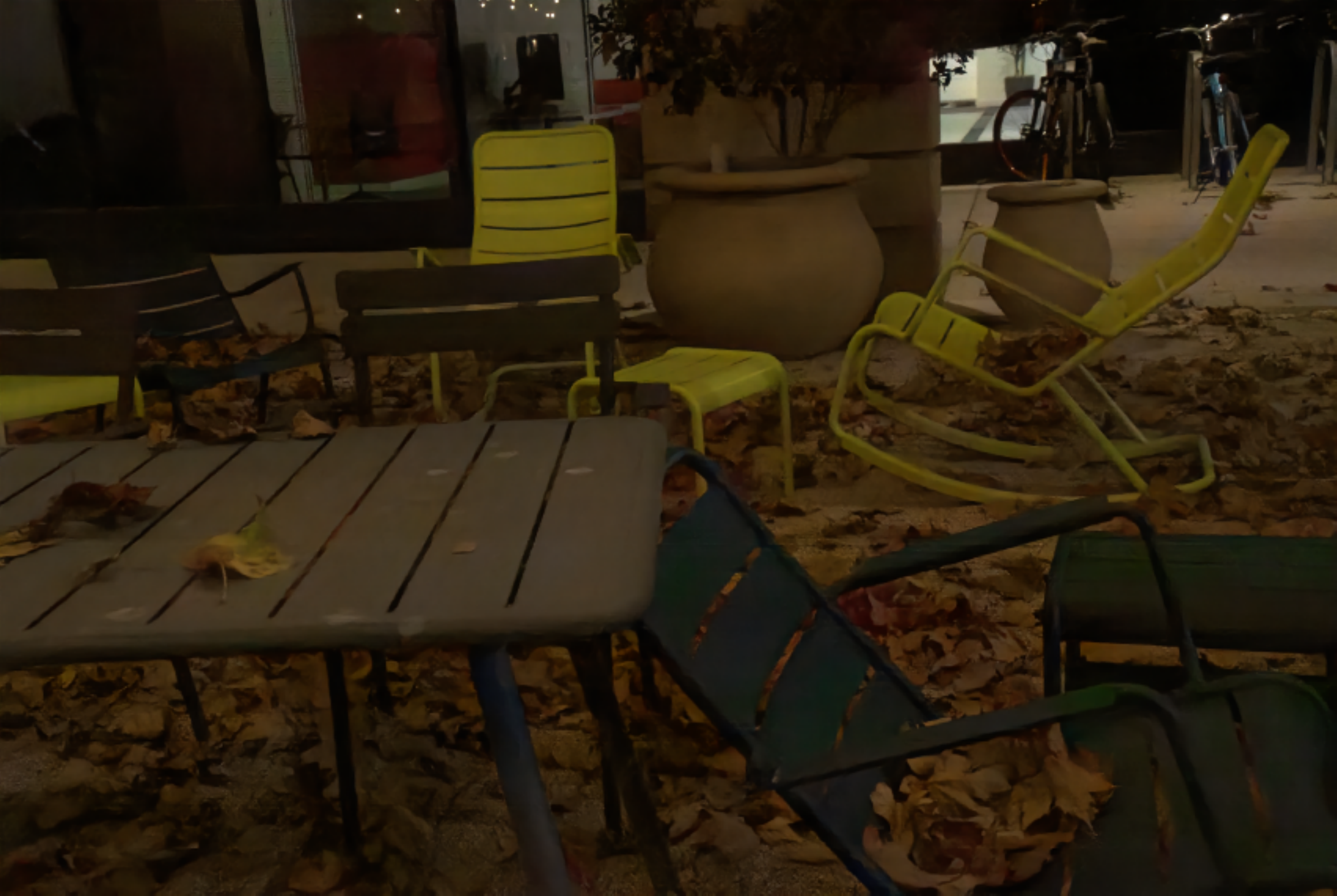}}
  \centerline{(d)}  
\end{minipage}

\caption{SID results of our method compared to original (without pruning). (a)(c) Original (PSNR: 28.54, SSIM: 0.767). (b)(d) Pruned with Method-D (PSNR: 28.55, SSIM:0.768)}
\label{fig:sidnodrop}
\end{figure}

\begin{figure}
\begin{minipage}{0.5\linewidth}
  \centering
  \centerline{\includegraphics[height=4.0cm]
  {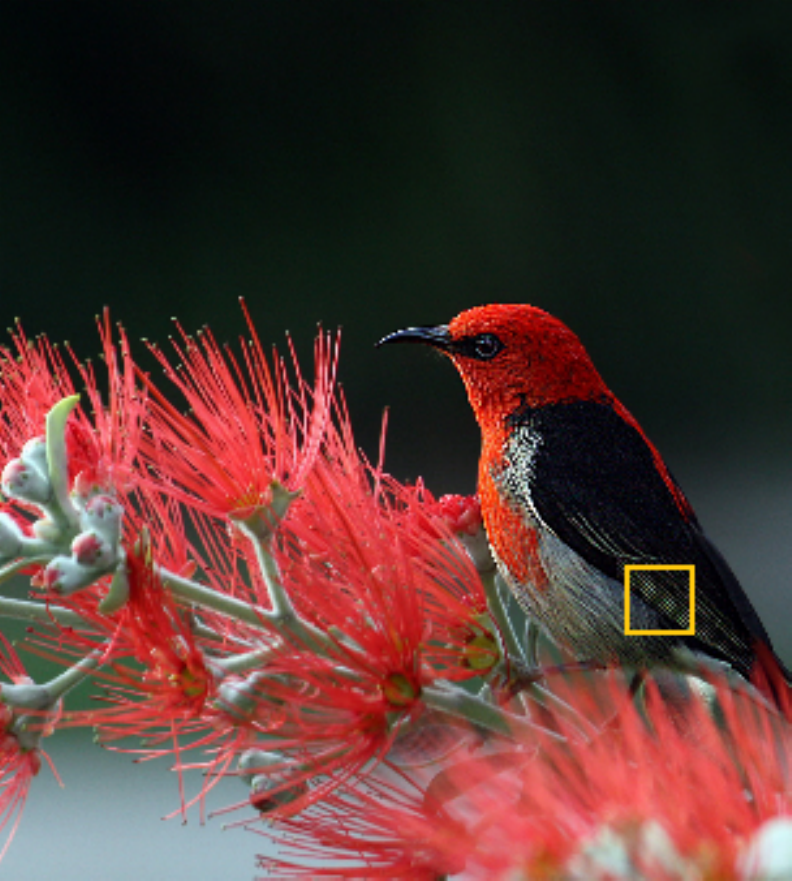}}
  \centerline{(a)}
\end{minipage}
\begin{minipage}{0.4\linewidth}
\begin{flushright}
	\begin{minipage}{1\linewidth}
  	\centering
  	\centerline{\includegraphics[height=1.6cm]
  	{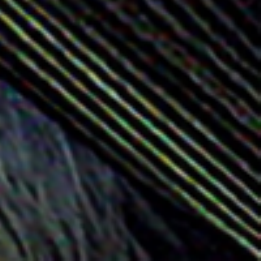}}
	\centerline{(b)}  	
  	\end{minipage}

	\begin{minipage}{1\linewidth}
  	\centering
  	\centerline{\includegraphics[height=1.6cm]
  	{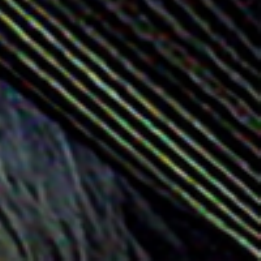}}
	\centerline{(c)}
	\end{minipage}
	\end{flushright}
\end{minipage}
\caption{EDSR results of our method compared to original (without pruning). (a) Image sampled from DIV2K. 
(b) Original (PSNR: 34.42, SSIM: 0.942). (c) Proposed Method-D (PSNR: 34.42, SSIM: 0.942)}
\label{fig:bird}
\end{figure}

\begin{figure}[htb]
\begin{minipage}{1\linewidth}
  \centering
  \centerline{\includegraphics[width=8cm]
  {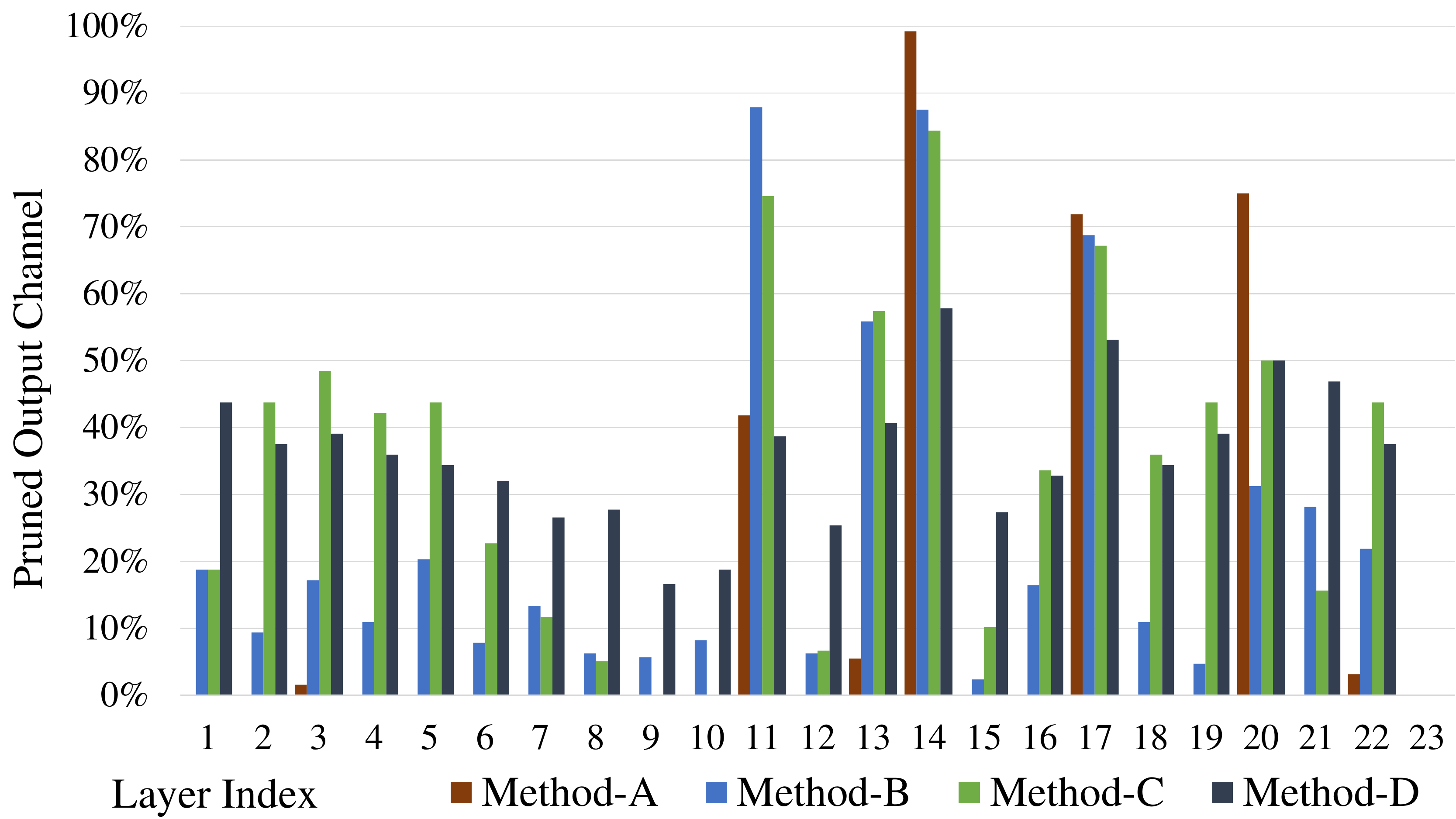}}
  \medskip
\end{minipage}
\caption{Pruned output channel per layer on SID}
\label{fig:sid_channel}
\end{figure}

\begin{figure}[htb]
\begin{minipage}{1\linewidth}
  \centering
  \centerline{\includegraphics[width=8cm]
  {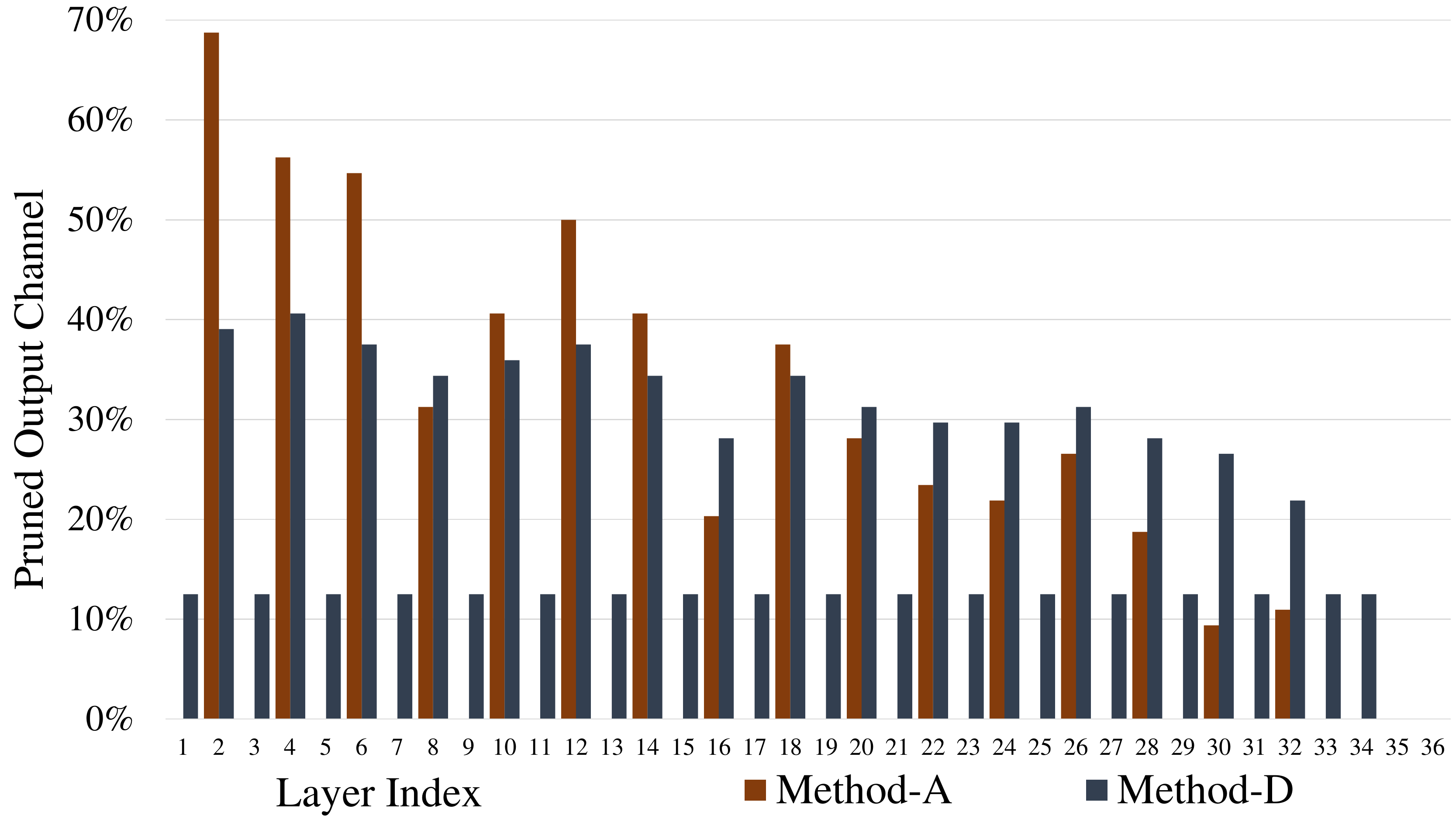}}
  \medskip
\end{minipage}
\caption{Pruned output channel per layer on EDSR}
\label{fig:edsr_channel}
\end{figure}

\subsection{Result}
The comprehensive analysis is elaborated in Table \ref{tab:expresult}.
We evaluate four distinct approaches.
Method-A stands for magnitude threshold pruning without any structural hints. Method-B keeps the depth of network (Sec. \ref{ssec:keepnetworktopology}). 
Method-C further considers MAC/weight ratio (Sec. \ref{ssec:macefficient}) on the basis of Method-B.
Method-D integrates all proposed techniques. All methods are conducted in company with quality-metric constraints (Sec. \ref{ssec:qualitymetricguarantee}).
For both SID and EDSR, it shows no PSNR and SSIM drop on all methods.
Fig. \ref{fig:sidnodrop} and Fig. \ref{fig:bird} reveal the indistinguishable quality difference.\\

{\setlength{\parindent}{0cm}\textbf{Keep Layer Depth.}
In SID, Method-B reduces MAC from 82\% to 63\% compared to Method-A. 
As shown in Fig. \ref{fig:sid_channel}, 
Method-A may remove all the output channels of a layer due to not keeping layer depth, which leads to severe quality metric drops that cannot be recovered in retraining steps.\\

{\setlength{\parindent}{0cm}\textbf{Enhance MAC Efficiency.}
In SID, Fig. \ref{fig:sid_channel} shows that Method-C prunes more weights on both top and bottom layers that have larger MAC/weight (Eq. \ref{equ:macefficient}).
Therefore, Method-C reduces MAC from 63\% to 48\% compared to Method-B.\\

{\setlength{\parindent}{0cm}\textbf{Balance Pruned Output Channel.}
Method-D increases 17\% weight sparsity but only reduces 6\% MAC compared to Method-C in SID. 
Fig. \ref{fig:sid_channel} illustrates that Method-D prunes less on top and bottom layers which have larger MAC/weight.\\

In EDSR, there is no difference among Method-A, Method-B and Method-C because no layer is pruned by Method-A 
and MAC/weight are identical for all layers (last layer could not be pruned) as shown in Fig. \ref{fig:macperweight}. 
Fig. \ref{fig:edsr_channel} shows that Method-D reduces MAC from 76\% to 63\%, which is more than 10\%, in shortcut-connected layers.\\

In summary, our methodology has significant reduction on both MAC and BW, which implies reduction on inference latency. 
For BW, we have 39\% and 20\% reduction on SID and EDSR, respectively. 
Our methodology also works well on complex network 
architecture.

\section{Conclusion}
To minimize computation complexity 
without quality drop on vision quality applications, 
our architecture-aware pruning 
is optimized for pruning more for complexity metric (e.g., MAC) on SID 
and shortcut-connected layers on EDSR. 
The MAC of SID and EDSR 
are reduced by 58\% and 37\%, respectively. 
Memory bandwidth is also reduced
without degradation of PSNR, SSIM and subjective quality. 
The reduction of computation complexity and memory bandwidth
could benefit on general mobile devices 
without special hardware design.

\bibliographystyle{IEEEbib}
\bibliography{reference}

\end{document}